\documentclass{svmult}

\usepackage{graphicx}

\begin{document}

\title{Multiwavelength Studies of Young OB Associations}

\author{Eric D. Feigelson}

\institute{Eric D. Feigelson \at Department of Astronomy \& Astrophysics, Pennsylvania State University, University Park PA 16802, \email{edf@astro.psu.edu}}

\maketitle

\abstract{We discuss how contemporary multiwavelength observations of young OB-dominated clusters address long-standing astrophysical questions: Do clusters form rapidly or slowly with an age spread? When do clusters expand and disperse to constitute the field star population? Do rich clusters form by amalgamation of smaller subclusters? What is the pattern and duration of cluster formation in massive star forming regions (MSFRs)?  Past observational difficulties in obtaining good stellar censuses of MSFRs have been alleviated in recent studies that combine X-ray and infrared surveys to obtain rich, though still incomplete, censuses of young stars in MSFRs.  We describe here one of these efforts, the MYStIX project, that produced a catalog of 31,784 probable members of 20 MSFRs.  We find that age spread within clusters are real in the sense that the stars in the core formed after the cluster halo.  Cluster expansion is seen in the ensemble of (sub)clusters, and older dispersing populations are found across MSFRs.  Direct evidence for subcluster merging is still unconvincing.  Long-lived, asynchronous star formation is pervasive across MSFRs.}

\section{Historical discussions of star cluster formation} \label{Feigelson_questions.sec}

Galactic Plane star clusters, well-known to classical astronomers like 2$^{nd}$ century Claudius Ptolemy and 10$^{th}$ century Abd al-Rahman al-Sufi,  were catalogued in the 18-19$^{th}$ centuries by Charles Messier and William and John Herschel.  As astrophysical explanations for astronomical phenomena rose to prominence around the turn of the 20$^{th}$ century, it was natural that the processes giving rise to clusters were investigated.  We address here several astrophysical themes of long-standing importance where, even today, theory is not well-constrained by observation.  

The historically oldest issue is the argument that most stars are born in clusters that expand and disperse to comprise the field star population.  In a 1917 discussion of Kapteyn's `systems of stars which travel together in parallel paths', Charlier \cite{Charlier17}, director of Lund Observatory in Sweden, argues
\begin{quote}
that the stars which now belong to such a system are only the insignificant remnant of a large cluster which at one time constituted a compact system in space.   
\end{quote}
Such questions could be investigated computationally, both by integrating difficult differential equations and by Monte Carlo N-body calculations, in the 1970s.  In an important study, Tutukov \cite{Tutukov78} wrote:    
\begin{quote}
It is generally believed that ... stars [form] in small groups which dissolve comparatively quickly during very early stages of evolution, practically at the moment of their formation. ... It is natural to suppose that the gas not utilized for star formation was blown away by hot stars, probably due to the ionizing radiation and stellar wind.  If the mass of gas is higher than the mass of stars and the kinetic energy of the gas exceeds the binding energy of the cluster, then the disruption of a young cluster seems inevitable.  
\end{quote}

The issue of stellar dispersal arose again when early-type stars were discovered far from their natal clouds away from the Galactic Plane. Greenstein \& Sargent \cite{Greenstein74} noted:
\begin{quote}
The kinematical behavior of these stars is, however, quite strange ... The stars are not kinematically relaxed; they are apparently observed soon after formation and ejection. ... [This reveals] a fundamental problem that far too many, hot, high-velocity, apparently normal stars exist.
\end{quote}
Some of these stars are clearly runaway stars ejected at high velocities from hard binary interactions, but others some dispersed up to $\sim 200$~pc from the Plane could not easily be traced to rich clusters \cite{deWit05}. In a catalogue of stellar members in OB associations within 3~kpc,  Garmany \& Stencel \cite{Garmany92} found that massive OB stars are commonly spread over large ($\sim 200$~pc) regions; these did not appear to be high-velocity runaways.  

Another long-standing issue concerns the mechanism by which rich star clusters form.   Aarseth \& Hills \cite{Aarseth72} sought to evaluate two alternatives views: simultaneous formation of a monolithic rich cluster and its possible later construction from pre-existing subclusters.  They wrote:
\begin{quote}
The density distribution of stars in a stellar cluster usually gives every appearance of being smoothing varying and non-clumpy.  On the face of it, this is a bit surprising since elementary considerations from [Jeans gravitational collapse] star-formation theory suggest that a cluster should initially be subdivided into a hierarchy of subclusters.  ... The subdivision process terminates when the cloud becomes opaque enough for the collapse time-scale to catch up with the cooling time-scale ... [so] that a cluster is initially composed of a hierarchy of subclusters.
\end{quote}
Stellar subgroups were empirically found in a number of nearby rich OB associations by Blaauw \cite{Blaauw64}.   But it was unclear whether the primary process is fragmentation of an initially homogeneous cluster, or incomplete consolidation of smaller subclusters into a unified structure.  The latter view came to the fore when molecular clouds were discovered to be highly inhomogeneous due to supersonic turbulence \cite{MacLow04}.  Maps obtained with the Herschel satellite far-infrared imaging show that even the coldest and densest cloud structures mostly have clumpy and filamentary structure \cite{Andre10}.  

A third contentious issue is the duration of star formation in molecular clouds.  Various researchers argue, on both physical and observational grounds, that cluster formation is rapid, although a small number of stars may form over an extended period before the principal starburst \cite{Elmegreen00, Palla00, Hartmann12}.   Others suggest that regulation of star formation by magnetically induced turbulence in molecular clouds and feedback from nascent stars prevents large-scale free-fall gravitational collapse and rapid cluster formation \cite{MacLow04, Bate09, Krumholz07, Krumholz12}.  The evidence outlined above for widely distributed early-type stars suggests that star formation in massive star forming regions are long-lived, so that earlier generation of massive stars have time to drift outward from still-active star forming regions. 

\section{The Observational Challenges} \label{Feigelson_challenges.sec}

It is now clear that most stars form in rich clusters. The cluster luminosity function in the Milky Way Galaxy and nearby galaxies demonstrates that the majority of stars form in clusters with $10^2-10^4$ stars \cite{Lada03} and, during galactic starburst episodes, superclusters of $10^5$ stars may dominate.  But even the fundamental physical properties, processes and timescales of cluster formation and early evolution are observationally poorly established.  Cogent arguments have been made that clusters form quickly \cite{Elmegreen00} and slowly \cite{Tan06}, that they form as a unified structure or are assembled from merging subclusters \cite{McMillan07, Bate09}, that they form in spherical cloud cores or in filamentary cloud structures \cite{Rathborne06, Andre10}.  Timescales for cluster formation and early dynamical evolution are poorly constrained by observation.  Attempts to measure the ages of constituent stars of nearby clusters by fitting their location in Hertzprung-Russell diagrams (HRDs) to theoretical evolutionary tracks is beset with observational difficulties, so that it is unclear whether the observed spreads in HRDs represent true age spreads \cite{Preibisch12}.  

The reasons for the failure to test competing astrophysical models of cluster formation can arguably be placed on practical observational difficulties in defining their member stars.  Much progress has been made in studying the progenitor molecular clouds through, for example, maps of coolant molecular lines with millimeter array telescopes and far-infrared imaging of continuum dust emission with the Herschel satellite.  The environmental effects of the hot OB stars can also be traced across the Galactic Plane: ionized gas is easily mapped at radio wavelengths, and heated dust produces PAH band emission mapped with infrared space telescopes.  But the actual stellar populations of star clusters beyond distances $\sim 1$~kpc are poorly known.  Indeed, hardly any members have been identified in most of the massive Galactic star forming regions that would be called `extragalactic giant H~II regions' were they to be present in nearby galaxies \cite{Figer08}.    

Acquiring a reliable census of members of star clusters beyond $d \sim 1$~kpc faces several challenges.  The most devastating is contamination by uninteresting older Galactic field stars along the line-of-sight.  At Galactic latitude $b \sim 0^\circ$ and longitudes in the inner quadrants, field stars have $10-100$ times higher surface density than the cluster members over most of the cluster extent at near-infrared magnitudes around the peak of the Initial Mass Function.  Interstellar absorption can reach $A_V \sim 30$~mag along the line-of-sight to the cluster, and can vary by tens of magnitude within the star forming region due to the local molecular cloud.  Detection of faint infrared stars is difficult amid the nebular H~II region emission from heated dust. 

As a result of these problems, the census of young star cluster members has often been restricted to nearby lower-mass clusters or to special subpopulations of massive clusters: the inner cluster core where the surface density rises above the field stars; OB stars that are brighter and bluer than ambient stars and easily confirmed with optical spectroscopy; and pre-main sequence stars with photometric infrared excesses (IRE) from dust protoplanetary disks.  The IRE criterion is often used to define the population of `young stellar objects' (YSOs) but it is restricted to disk-bearing pre-main sequence stars (Class I-II).  In many clusters, the bulk of the stars have lost their disks and are thus photometrically indistinguishable from contaminant field stars in the infrared bands.  Inferences regarding star formation histories may be flawed due to the IRE sample bias towards younger systems with hot inner accretion disks. 

However, a technique has emerged in recent years that overcomes, to some degree, these observational difficulties and biases.  Sensitive and high-resolution imaging of star forming regions with NASA's Chandra X-ray Observatory, sensitive in the $0.5-8$~keV ($25-1.5$~\AA) X-ray band, can detect reasonable fractions of young cluster populations out to distances of several kiloparsecs with reasonable exposure times.   A typical 100~ksec exposure with Chandra's Advanced CCD Imaging Spectrometer of a typical rich cluster at $d \sim 2-3$~kpc will reveal 1000 or more cluster members, perhaps $5-20$\% of the full Initial Mass function (IMF).  Most importantly, the X-ray image captures only a minute fraction of the Galactic field stars that contaminate the infrared images so badly.  The main contaminant of X-ray images are quasars seen through the Galactic Plane, and these are readily removed due to their lacking infrared counterparts.  X-ray emission in pre-main sequence arises from magnetic flaring activity, similar to that of the Sun but with much more powerful and frequent flares \cite{Feigelson99}.  The flaring X-ray emission has a sufficiently `hard' X-ray spectrum that these stars can be detected through high column densities of intervening interstellar material, equivalent to $A_V \sim 100$ mag in some cases.  Finally, X-ray selection is complementary to IRE selection because it most efficiently captures disk-free (Class III) stars.  

The remainder of this chapter discusses a particular effort called MYStIX (Massive Young Stellar complexes study in Infrared and X-rays) that combines Chandra X-ray, UKIRT near-infrared, and Spitzer Space Telescope mid-infrared surveys of 20 OB-dominated star forming regions at distances $0.4 < d < 4$~kpc \cite{Feigelson13}.  After complicated data analysis with statistical procedures designed to reduce contaminants, a sample of $\sim$31,000 MYStIX Probable Complex Members (MPCMs) is generated.  While far from a complete stellar census, the samples are typically much larger than previously available, and appear to be reasonably free from contaminating field stars.  After a brief description of the MYStIX observational effort (\S\ref{Feigelson_MYSTIX.sec}) and a new stellar chronometer based on X-ray/infrared photometry (\S\ref{Feigelson_chron.sec}), we summarize some of the characteristics of these star clusters (\S\ref{Feigelson_subclusters.sec}).  A variety of results are then outlined (\S\ref{Feigelson_results.sec}): the morphology of stellar clustering and maps of stellar surface density, histories of star formation in MSFRs, and direct measurement of cluster expansion.  

MYStIX is only one of several similar X-ray/infrared surveys that include: Chandra Carina Complex Project \cite{Townsley11}, Chandra Cyg OB2 Legacy Survey \cite{Wright14b},  Star Formation in Nearby Clouds \cite{Getman17}, NGC~6611 \cite{Guarcello07},  Eagle Nebula \cite{Guarcello10}, NGC~1893 \cite{Prisinzano11}, DR 15 \cite{Rivera15}, NGC 6231 \cite{Damiani16, Kuhn17}, NGC 7538 \cite{Sharma17}, and others.

\section{The MYStIX Project} \label{Feigelson_MYSTIX.sec}

The MYStIX effort  seeks to construct an improved census of stars in rich clusters and their environs in 20 MSFRs near the Sun.  Populations that are not dominated by an O or early-B star are omitted; thus MYStIX omits nearby small star forming regions like the Taurus-Auriga, $\rho$ Ophiuchi and Chamaeleon complexes.  Table~1 lists the MYStIX star forming regions with approximate distance from the Sun and spectral type of the dominant star.  The accompanying Figure 1 shows the location of the MYStIX regions on a diagram of the Milky Way Galaxy with the Sun at the middle.   The MYStIX targets do not constitute a complete sample in any way, but rather were selected by practical considerations: they must have sufficiently deep coverage by the Chandra and Spitzer satellite imagers.  

{\vspace{10pt}\hspace*{-20pt
\begin{minipage}{0.6\textwidth}
\noindent {\bf Table 1} MYStIX Star Forming Regions  \\~\\
\begin{tabular}{| lll | llll |} \hline 
Region & D$_{kpc}$ & ~* 	&& 		Region & D$_{kpc}$ & ~* \\ \hline
Orion Neb		& 0.4 & O7 	&&		NGC 6334		& 1.7 & O8: \\	
Flame Neb		& 0.4 & O8:	&&		NGC 6357		& 1.7 & O3 \\
W 40			& 0.5 & O:  	&&		Eagle Neb		& 1.8 & O9 \\
RCW 36			& 0.7 & O8 	&&		M 17			& 2.0 & O4 \\
NGC 2264		& 0.9 & O7 	&&		W 3			& 2.0 & O5 \\
Rosette Neb		& 1.3 & O4 	&&		W 4			& 2.0 & ~... \\
Lagoon Neb		& 1.3 & O4 	&&		Carina Neb	& 2.3 & O2 \\
NGC 2362		& 1.5 & O9I	&&		Trifid Neb		& 2.7 & O7 \\
DR 21			& 1.5 & ~... 	&&		NGC 3576		& 2.8 & O: \\
RCW 38			& 1.7 & O5 	&&		NGC 1893		& 3.6 & O5 \\ \hline 
\end{tabular}
\end{minipage} ~
\begin{minipage}{0.42\textwidth}\vspace{10pt}
\includegraphics[width=\textwidth]{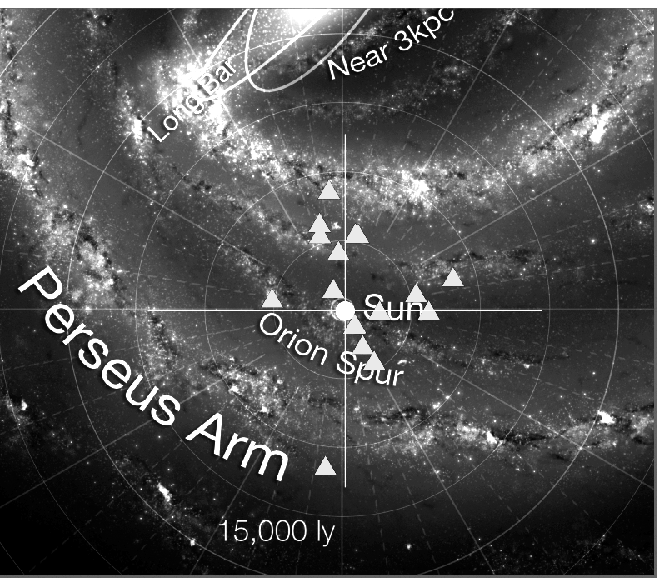}
{\small Figure 1. Galactic location of MYStIX star forming regions (triangles)}
\setcounter{figure}{1}
\end{minipage}
\vspace{10pt}
}

Simply stated, the MPCM samples are the sum of probable complex members extracted from X-ray sources in the Chandra X-ray Observatory images, IRE sources from UKIRT near-infrared observations (often part of the UKIDSS Galactic Plane Survey) and the Spitzer Space Telescope mid-infrared observations, and published OB stars confirmed by published optical spectroscopy.  But the actual procedure for constructing the MPCM samples is complicated by the need to reduce the often-overwhelming contamination of Galactic field stars combined with spatially variable cloud absorption and nebular emission.  Challenges overcome include: 
\begin{description}

\item[\bf X-ray source lists] \hspace*{-10pt} were obtained using the {\it ACIS Extract} package and associated software developed for the Chandra ACIS instrument at Penn State \cite{Kuhn13a, Townsley14}.  This allows detection of sources with as few as $3-5$ photons on-axis, even in the presence of crowding and diffuse X-ray emission.  Contamination from extragalactic X-ray sources and field X-ray stars was reduced by a naive Bayes classifier based on various properties of the sources and their infrared counterparts \cite{Broos13}. The reliability of these sources is validated by the high fraction associated with stars exhibiting other pre-main sequence properties \cite{Kuhn13a, Kuhn17}. 

\item[\bf Near-infrared source lists] \hspace*{-10pt} were obtained with the UKIDSS pipeline software modified to accommodate very crowded Galactic plane fields with nebulosity \cite{King13}.

\item[\bf Mid-infrared source lists] \hspace*{-10pt} were obtained with the Spitzer IRAC team software modified to accommodate crowding and nebulosity \cite{Kuhn13b}.

\item[\bf X-ray/infrared counterpart identifications] \hspace*{-10pt} were based on a probabilistic calculation of proximate sources that accounts for the magnitude distribution expected for true complex members, in order to reduce false associations with fainter field stars \cite{Naylor13}.

\item[\bf Infrared excess stars] \hspace*{-10pt} were extracted based on a complicated decision tree of criteria designed to reduce the often-heavy contamination by field red giants and false sources associated with nebular knots \cite{Povich13}.

\end{description}

The classified X-ray sources, IRE stars and published OB stars were then combined into the MPCM catalog of 31,784 stars in the 20 regions of Table~1 \cite{Broos13}.  The MYStIX papers, and their electronic tables of intermediate and final samples, are collected at the Web site http://astro.psu.edu/mystix.  

The MPCM sample is far from a complete census.  The X-ray samples are generally limited to stars with masses above $\sim 0.5$~M$_\odot$, and thus miss the peak of the IMF of low-mass members.  Various biases are present in the sample as well (see Appendix B of Feigelson et al. 2013).  Nonetheless, the MPCM samples are the largest for most of the star forming regions under consideration.  Tests of the sample reliability were made using the well-studied NGC~2264 population; $\sim 80$\% of previously identified H$\alpha$ and optically variable stars were recovered, and dozens of new members are proposed \cite{Feigelson13}.  

\begin{figure}
\centering
\includegraphics[height=0.37\textheight]{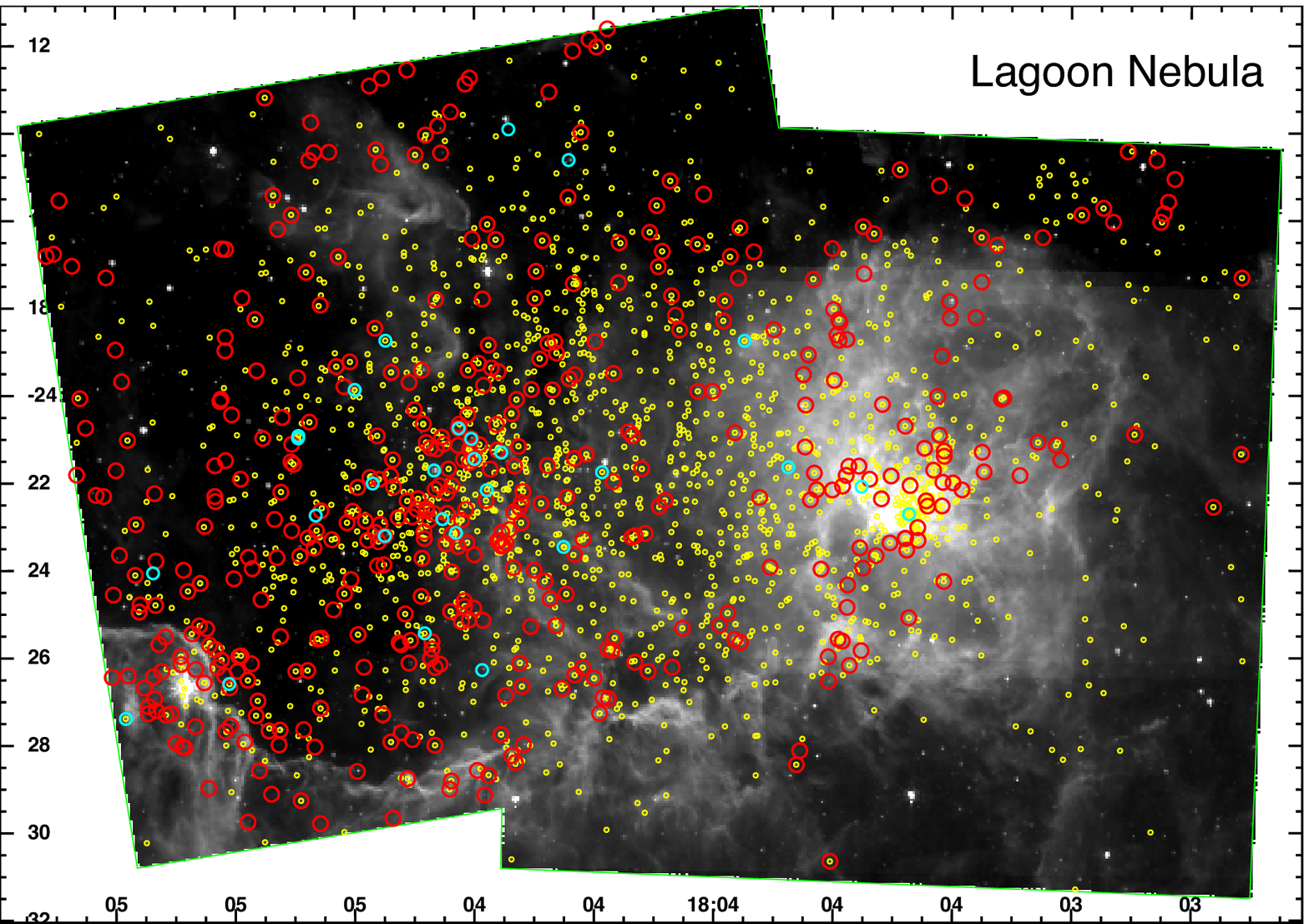} \\ \vspace{10pt}
\includegraphics[height=0.48\textheight]{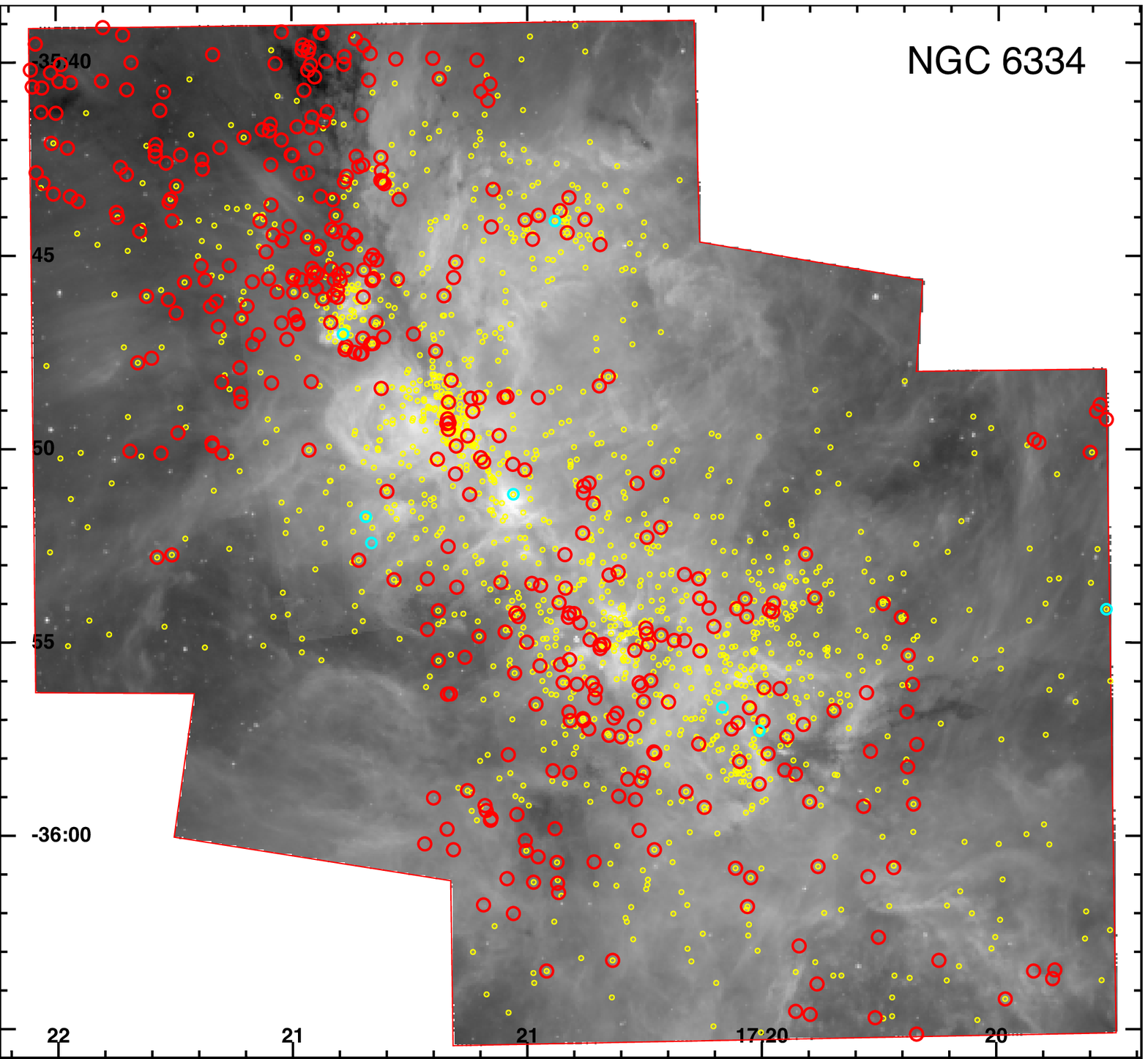} 
\caption{Spatial location of MYStIX Probable Complex Members (MPCMs) for the Lagoon Nebula and NGC~6334 fields \cite{Broos13}.  Infrared excess stars are noted by red circles, X-ray selected stars by yellow dots, and published OB stars by cyan circles.  The stars are superposed on Spitzer IRAC 8.0$\mu$m maps.  Each Chandra field subtends $17^\prime \times 17^\prime$.  \label{mpcm.fig}}
\end{figure}

Figures~\ref{mpcm.fig}-\ref{MPCM.fig} illustrate the MPCM samples for four MYStIX star forming regions.  The regions have complex structures though with some similar behaviors.
\begin{description}

\item[\bf Lagoon Nebula (M 8)] In this MSFR,  we see two major clusters: the poorly characterized NGC~ 6523 cluster to the east with the famous massive star Herschel~36; and the well characterized NGC~6530 cluster in a large cavity to the west.  As one proceeds westward, the fraction of IRE stars (red circles in Figure~\ref{mpcm.fig}) decreases; it is not immediately clear whether this is an age gradient or a selection effect due to the difficulty of finding IRE stars in the bright PAH nebulosity of the western region.  A clump of stars is also seen to the far-southeast associated with a bright rimmed cloud; it includes the luminous embedded star M 8E.   

\item[\bf NGC~6334] This is a large MSFR elongated along the Galactic Plane with both heavy absorption and complex bright nebular emission that precluded generation of a reliable stellar census in the past.  The 1,667-member MPCM sample shows several distinct clusters, some dominated by young IRE stars and others by older X-ray selected stars \cite{Feigelson09}.  The morphology might represent a star formation wave from the southwest to the northeast, but older clusters are sometimes superposed on younger clusters and a distributed young star component is also present.  A selection of likely protostars, based on MYStIX sources with ascending infrared spectral slopes or ultra-hard X-ray spectra, shows a distribution of very young stars tracing the curved molecular filament to the northeast \cite{Romine16}. 

\item[\bf NGC~6357]  This region has 2,235 MPCMs, very few of which had previously been identified by optical or infrared surveys even though this is a very active star forming region in the Carina spiral arm.  Three very rich clusters are seen; Pismis 24 to the northwest has several $\sim 100$~M$_\odot$ O3 stars.  In each cluster, we can see spatial displacements between the infrared and X-ray selected subsamples.  The IRE selection method is ineffective around the brightest nebular emission of the northwest H~II region.  Two dozen new absorbed ($4 < A_V < 24$ mag) candidate OB stars are identified in the MYStIX catalog in this region \cite{Povich17}. 

\item[\bf Eagle Nebula (M 16)]  Here the southwestern rich cluster is dominated by disk-free X-ray selected members, while the sparser subclusters to the north and west are dominated by disk-bearing IRE members.  As in most MYStIX regions, the X-ray selected stars outnumber the IRE stars, implying that the star formation has endured for many millions of years beyond the typical longevity of infrared-emitting disks.   

\end{description}

\begin{figure}[t]
\includegraphics[width=1.0\textwidth]{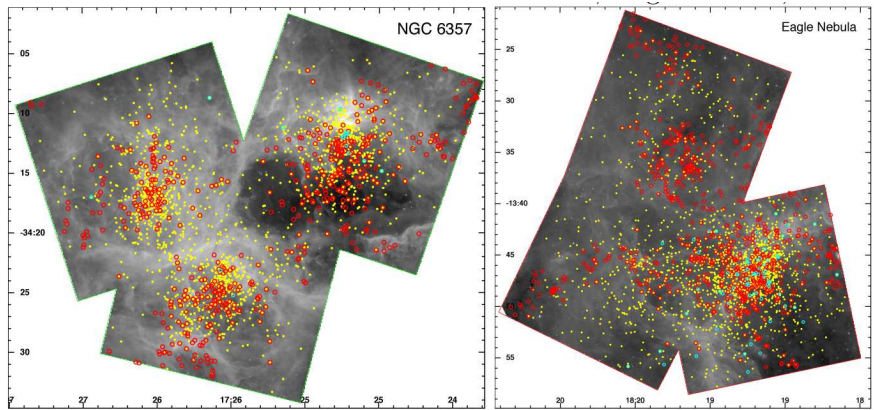}
\caption{The MPCMs in the NGC~6357 and Eagle Nebula complexes superposed on Spitzer  IRAC 8$\mu$m maps \cite{Broos13}.  Yellow dots are X-ray selected members, red circles are infrared-excess members, and cyan symbols are published OB stars. \label{MPCM.fig} \vspace*{-15pt}}
\end{figure}

\begin{figure}[t]
\includegraphics[width=0.5\textwidth]{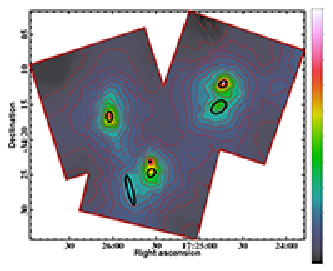}
\includegraphics[width=0.5\textwidth]{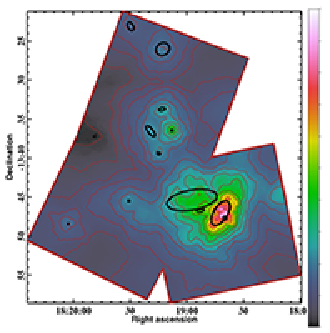}
\caption{Statistically defined subclusters in the NGC~6357 and Eagle Nebula complex are shown as black ellipses superposed on smoothed maps of the MPCM stellar distribution. \label{subclusters.fig}}
\end{figure}

\section{A New Stellar Chronometer} \label{Feigelson_chron.sec}

To reveal the spatio-temporal history of star formation in MYStIX regions, it would be very desirable to obtain reliable ages of different (sub)clusters of MPCM stars.  Two pre-main sequence chronometers are traditionally used: a star's location in the HRD compared to theoretical evolutionary tracks; and the presence of a star's infrared-emitting circumstellar disk \cite{Haisch01, Richert17}.  But neither are very effective for MSFRs.  HRD locations are not available because the stars are often too reddened to readily obtain optical spectra, and in any case several extraneous problems render HRD-derived ages uncertain \cite{Preibisch12}.  Disk fractions or classification (Class 0-I-II-III) derived from infrared photometry are inaccurate and difficult to calibrate.  For example, IRE populations are reduced by local H~II region contamination, differences in infrared-to-X-ray sensitivities can systematically bias disk fraction comparisons between MYStIX regions, and individual disk dissipation timescales range over $0.5-5$~Myr or more.  A potentially accurate chronometer based on oscillations of intermediate-mass stars has been proposed \cite{Zwintz14}, but it can be applied only to a handful of bright well-studied stars, not to thousands of faint MSFR stars. 

In the MYStIX context, Getman and colleagues have developed a new, surprisingly simple chronometer for pre-main sequence stars that can be applied to a reasonable fraction of MPCM stars \cite{Getman14a}.  It is based on the long-standing empirical correlation between X-ray luminosity $L_x$, produced by magnetic reconnection flares, and stellar mass $M$ in pre-main sequence stars.  This $L_x-M$ relation is best calibrated in the Taurus-Auriga population \cite{Telleschi07}.  The astrophysical cause of this correlation is poorly understood (presumably related to magnetic dynamos in fully convective stellar interiors), but it accounts for much of the $10^4$ range of $L_x$ in young stellar populations.  MYStIX $L_x$ measurements, after correction for soft X-ray absorption from intervening interstellar gas, thus give mass estimates for each star.  MYStIX also gives measures photospheric luminosities $L_{bol}$; Getman et al.\ use dereddened $J$ band magnitudes $M_J$ as a proxy for $L_{bol}$.   $M$ values inferred from $L_x$ and measured $M_J$ values combined with standard theoretical evolutionary tracks give stellar age estimates for each star, nicknamed $Age_{JX}$.  Each $Age_{JX}$ value is quite inaccurate, but obtaining the median $Age_{JX}$ for a spatially defined subsample of young stars appears to be effective for elucidating histories of star formation within and between clusters.    

\section{Identifying (Sub)Clusters } \label{Feigelson_subclusters.sec}

The MYStIX fields are mostly centered on rich OB associations with optically bright H~II region, often with names like `Rosette Nebula' and `Lagoon Nebula' that date to the 19$^{th}$ century.  But examination of the MPCM spatial distributions show considerable diversity in clustering behavior $-$ a simple dichotomy between rich clusters and distributed star formation is clearly inadequate.   Global statistics of spatial point processes, such as Ripley's $K$ function and the related two-point correlation function \cite{Illian08}, are not directly useful as they are strongly affected by the richest clusters and do not reflect the diversity of patterns within a single field.  Defining stellar `clusters' or `groups' by surface density enhancements \cite{Feigelson11} also has the disadvantage of requiring an arbitrary threshold.  

We therefore proceeded to locate `clusters' using a parametric statistical regression approach known as `mixture models' \cite{McLachlan00}.   Here we require that cluster structure have a specific mathematical form corresponding an isothermal sphere or ellipsoid \cite{Kuhn14a}.  A likelihood function giving the probability that the observed celestial locations of MPCM stars corresponds to a specified mixture of isothermal ellipsoids.  When a flat `distributed' stellar population is added, a model with $k$ clusters has $6k+1$ parameters.  The best fit model is obtained by maximum likelihood estimation for a range of $k$, and the optimal number of clusters is obtained by maximizing the Akaike Information Criterion, a well-accepted penalized likelihood measure for model selection. Note that the method permits hierarchical structures with one ellipsoid lying within or overlapping another ellipsoid. Model fits are generally excellent with no strong features in the residual spatial maps.   The resulting spatial decompositions for the NGC~6357 and Eagle Nebula MYStIX fields are shown in Figure~\ref{subclusters.fig}.  
 
The result of this analysis is the assignment of each of the $\sim$31,000 MPCM stars to one of 142 (sub)clusters or to a distributed population \cite{Kuhn14a}. Since each subcluster has an assumed isothermal ellipsoid internal structure, parameters such as core radii and ellipticity can be calculated.  Two measures of absorption are available for each (sub)cluster: the median $J-H$ color index and the sample median of the individual median energies of the X-rays from  the constituent stars.   For example, the Eagle Nebula has 12 statistically significant subclusters (Fig.~\ref{subclusters.fig}) with sample populations ranging from 7 to 451 MPCM stars, core radii from 0.07~pc to  1.0~pc, ellipticities from 7\% to 64\%, and absorptions from $A_V \sim 5$ to 16~mag.  Note that the sample populations are not unbiased measures of the true stellar populations, as they depend on the  circumstantial exposure times of the Chandra and Spitzer observations, region distance and absorption.

Two additional critical properties of subclusters can be derived. First, the age of each subcluster can be estimated from the median $Age_{JX}$ values of the constituent stars (Sec.~\ref{Feigelson_chron.sec}).  Ages for the Eagle (sub)clusters range from 0.8 to 2.4~Myr.  Second,  the total stellar population can be inferred  by scaling the sample X-ray luminosity function (truncated at different limiting X-ray sensitivities) to the full-sampled X-ray luminosity function of the Orion Nebula Cluster \cite{Kuhn15a}.  The total populations inferred from X-ray luminosity functions agree well with a parallel analysis based on dereddened $J$ band magnitudes scaled to a standard Initial Mass Function.  

Combining the estimated total population with the (sub)cluster structural parameters like core radius, unbiased estimates can be made of important quantities such as total stellar mass (in M$_\odot$), central surface densities (in stars/pc$^2$), central volume densities (in stars/pc$^3$), characteristic crossing and relaxation times (in Myr) \cite{Kuhn15b}.  

\section{Spatial distribution of stars across star forming regions} \label{Feigelson_maps.sec}

Comparisons of MPCM stellar spatial distributions  in maps like Figs.~\ref{MPCM.fig}-\ref{subclusters.fig} can be misleading due to inhomogeneity in sensitivity. This particularly affects the X-ray measurements.  First, within each Chandra ACIS field the sensitivity is highest at the field center and degrades by a factor of $\sim 3$ as one approaches the field edges due to the coma of the high-resolution X-ray mirrors.  Second, Chandra fields are often mosaics of overlapping exposures; due to the low background of the ACIS detector, sensitivity scales linearly with exposure time.  Third, the Chandra exposure times are not scaled with the square of the MYStIX region distance, so the X-ray luminosity function (and, through the empirical $L_x$-Mass relationship) and mass function are truncated at different levels. 

\begin{figure}
\includegraphics[width=1.0\textwidth]{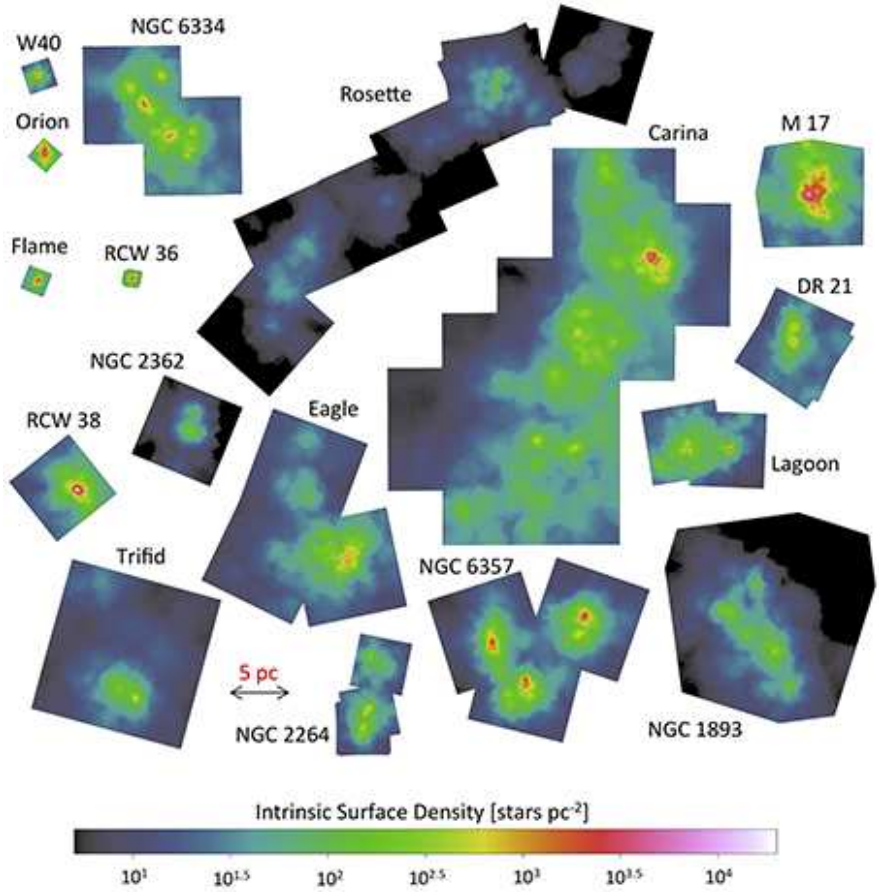}
\caption{Montage of maps of surface density of X-ray selected stars in MYStIX regions, shown to the same physical scale in parsecs \cite{Kuhn15a}. These intrinsic surface densities have been corrected for variations of X-ray sensitivities within and between fields, with density values scaled to the full IMF.  \label{mystix_surfdens.fig}}
\end{figure}

However, as outlined in Sec.~\ref{Feigelson_subclusters.sec}, these problems can be overcome \cite{Kuhn15a}.  We first `flatten' the intra-ACIS sensitivity variation by omitting the faint sources near the field center.  The stellar surface densities are then normalized to the full IMF assuming all regions have the same intrinsic X-ray luminosity function.  Although the lower mass stars missed by Chandra cannot be individually identified, the surface densities can be scaled upward to compensate for the different truncation levels.  Note it is more difficult to corrected the maps for variations in the surface densities of IRE sources, which are deficient in the brightest H~II nebular regions.  

The result is Fig.~\ref{mystix_surfdens.fig} a remarkable new view of the stellar distributions in massive star forming clouds  \cite{Kuhn15a}. The densities correspond to the full intrinsic stellar populations down to the M$\sim 0.08$~M$_\odot$ limit  shown on a uniform physical scale (see the 5~pc scale bar) and a uniform color scale in stars/pc$^2$ (see color calibration bar).  We find, for example, that the both the embedded clusters and the revealed massive cluster of the Rosette Nebula region have low surface densities of $10^1$ stars/pc$^2$. But the RCW~38, Orion Nebula Cluster, and M~17 clusters have extremely high central surface densities around $10^4$ stars/pc$^2$.  

Diversity, rather than consistency, is the premier result from these surface density maps.  The main Rosette Nebula cluster NGC~2244 must be in a completely different dynamical state than the RCW~38 or W~40 clusters; and indeed this may be related to the complete absence of mass segregation in NGC~2244 \cite{Wang08}.   Until these maps were compared, it was not realized that RCW~38 (which is badly contaminated in the IR bands due to nebulosity) has the densest collection of stars of any cluster in the nearby Galaxy.  It thus provides an excellent laboratory to study dynamical effects of close stellar encounters \cite{Pflamm06, Pfalzner13}.  

The MYStIX maps showing of a wide range of central surface densities, $<10^1$ to $\sim 3 \times 10^4$ stars/pc$^2$ (Fig.~\ref{mystix_surfdens.fig}), stands in conflict with the findings of Bressert and colleagues who report that young stellar clusters exhibit a characteristic central surface density distribution with mean around 20~stars/pc$^2$ \cite{Bressert10}.  Their study is limited to nearby molecular clouds where clusters are generally small and, most importantly, their sample is limited to IRE stars and thus miss the  disk-free X-ray selected stars that dominate many star forming regions.  The MYStIX findings on stellar surface densities, although still subject to limitations and biases, are probably more reliable than the more constrained IRE-only results.  

\section{Observational constraints on astrophysical questions} \label{Feigelson_results.sec}

We now discuss how the MYStIX project -- specifically the MPCM sample of 31,747 young stars in 142 (sub)clusters associated with 20 MSFRs -- addresses the astrophysical questions outlined in \S\ref{Feigelson_questions.sec} that concern the origin and early evolution of star clusters.   The questions are pursued by searching for spatial and statistical patterns among the various physical quantities measured or inferred for the (sub)clusters.  One must recognize that the MPCM sample is constructed in complicated ways with unavoidable incompleteness and biases \cite{Feigelson13}; however, each MYStIX region is analyzed in the same fashion and corrections to alleviate sensitivity and contamination effects can be applied in consistent ways.  

\subsection{Cluster expansion and dispersal}

The MYStIX dataset shows many cases of the expected range of cluster structures: compact clusters embedded in their molecular cores, larger clusters following molecular gas ejection, and older stars dispersing into the field population.  

Direct evidence for cluster expansion is shown in Fig.~\ref{expansion.fig} \cite{Kuhn14a, Kuhn15b}.  The first panel shows that MYStIX (sub)cluster core radii systematically increase as clusters range from heavily absorbed to lightly absorbed.  The X-ray median energy range is roughly equivalent to $0 < A_V < 40$; the same result is seen using $J-H$ as an absorption measure.  The other panels show show the relationships between core radii or central density and median $Age_{JX}$ values for the subclusters.  Here we see roughly a factor of 10 increase in radius, and a factor of 1000 decrease in central core density, as (sub)clusters age from $\sim 0.5$ to 4~Myr.  This is roughly consistent with dynamical calculations of cluster expansion following gas expulsion, although some models assume initial conditions that predict more rapid expansion at earlier times \cite{Banerjee13, Pfalzner13}. 

\begin{figure}
\centering
\includegraphics[width=0.37\textwidth]{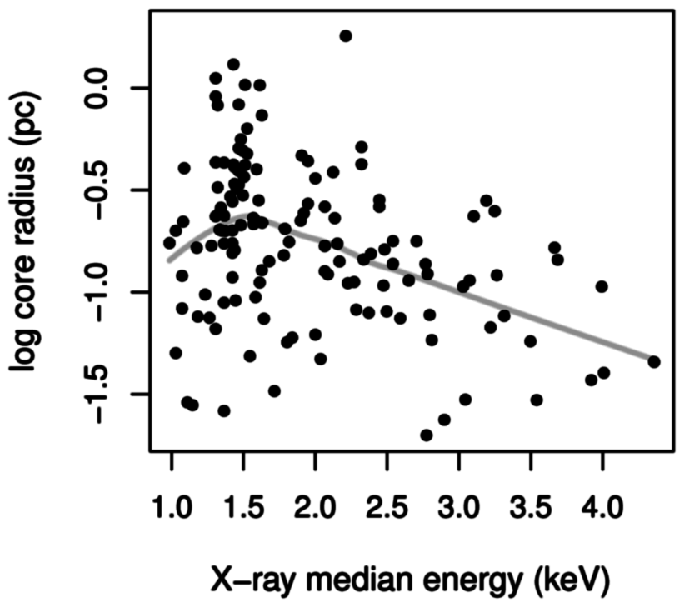} 
\includegraphics[width=0.62\textwidth]{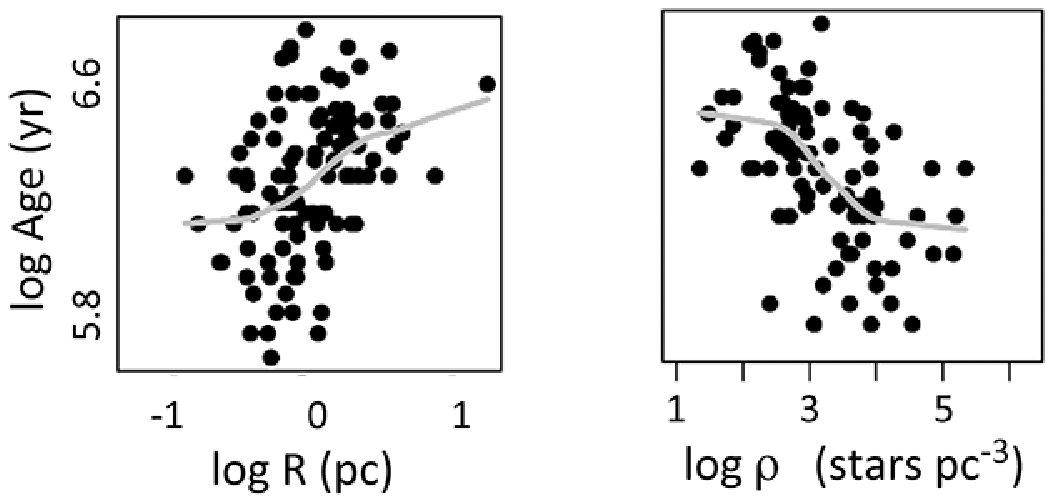}  
\caption{Expansion of young clusters.  {\it Left:} Subcluster core radius $vs.$ X-ray median energy, a measure of interstellar absorption with nonparametric regression curve \cite{Kuhn14a}.  {\it Center and right:} Bivariate scatter diagrams showing subcluster radius and central density $vs.$ $Age_{JX}$ with nonparametric regression curve \cite{Kuhn15b}.  \label{expansion.fig}}
\end{figure}

Evidence of this expansion have been presented by Pfalzner and colleagues \cite{Pfalzner09, Pfalzner13} using samples of Galactic and extragalactic young clusters obtained from the literature.  They report a `universal sequence' relating cluster size, central density and age indicative of cluster expansion from a uniform compact state.  Their `loose clusters', similar to MYStIX clusters, expand $\sim 10$-fold from $2-20$~Myr.  

Pre-MYStIX studies had reported that X-ray selected stars, including early-type OB stars, were often dispersed from the molecular cores that active form stars today \cite{Feigelson09}.   In the Carina complex, half of the X-ray stars lie outside the regions dominated by the Trumpler 14-15-16 clusters and the South Pillars clouds \cite{Feigelson11}.  This pattern is seen in most MYStIX regions.  Dispersed stellar surface densities range from near-zero to tens of stars/pc$^2$  in the different regions \cite{Kuhn14a, Kuhn15b}. $Age_{JX}$ analysis shows that, in nearly all cases, the dispersed stars are older (typically 3 to $>$5~Myr) than the MYStIX (sub)clusters \cite{Getman14a}.  

These findings give confidence in the long-standing argument \cite{Charlier17} that young clusters often quickly dissipate to constitute the field star population.  However, the MYStIX photometric observations cannot distinguish the physical process: do individual stars slowly drift away, are individual stars ejected at high velocity by stellar interactions in the cluster core, or do clusters release all of their stars simultaneously as they become gravitationally unbound?

\subsection{Cluster formation by merging subclusters}

The MYStIX (sub)cluster sample gives ample opportunity to reveal merging of smaller subclusters as an important process of building up large equilibrated clusters as predicted in models of cluster formation in turbulent molecular clouds \cite{Bate09}.  Yet the evidence is unclear.

First, consider the geometric properties of MYStIX (sub)clusters without inclusion of physical quantities such as age and mass \cite{Kuhn14a}.  As exemplified in NGC~6357 and the Eagle Nebula decompositions in Fig.~\ref{subclusters.fig}, some rich clusters are consistent with simple smooth ellipsoidal stellar distributions, while others are clumpy and require several ellipsoids to be adequately modeled.  Fig.~\ref{subcluster_morph.fig} is a diagram of the ellipsoidal structures in 15 MYStIX regions placed into a heuristic classification of simple, linear chain, core-halo, and complex clumpy classes \cite{Kuhn14a}.  As in \S\ref{Feigelson_maps.sec}, we see a wide diversity of clustering morphologies produced by massive molecular clouds.  

\begin{figure}
\centering
\includegraphics[width=0.85\textwidth]{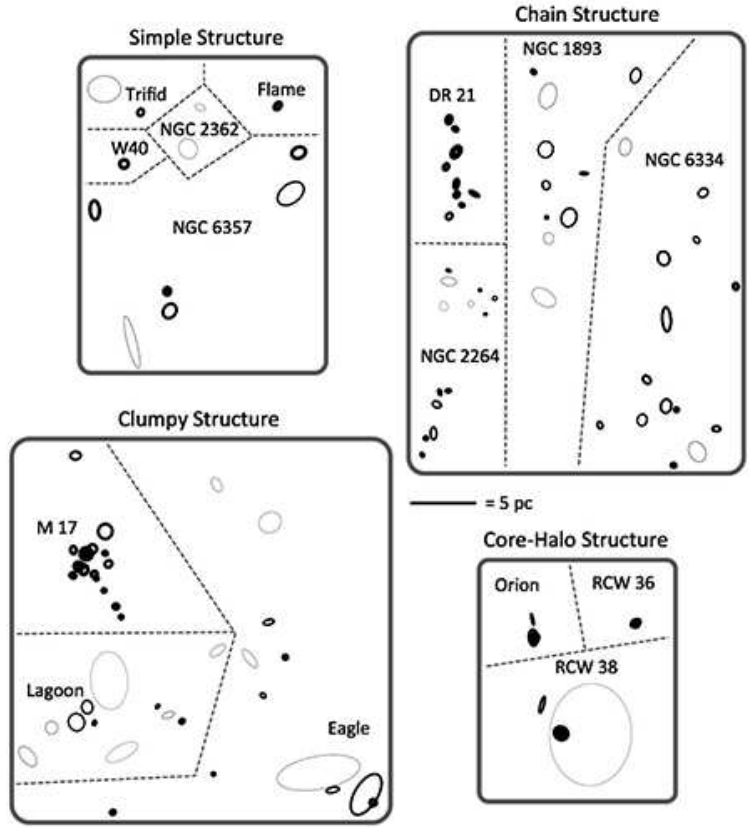}
\caption{Heuristic classification of the star formation complex morphology of 15 MYStIX regions based on the ellisoidal subcluster spatial decomposition \cite{Kuhn14a}.  Regions are shown on the same physical scale (see the 5~pc calibration bar), and line thickness is scaled to the subcluster stellar surface density of each subcluster. \label{subcluster_morph.fig}}
\end{figure}

It is tempting to interpret the morphological classes as an evolutionary sequence where   star formation begins as linear chains in filamentary clouds, passes through a clumpy stage as subclusters merge, and ends with core-halo and simple structures that may be in dynamical equilibrium.  However, when $Age_{JX}$ values are examined for these morphological classes, no evidence for an evolutionary sequence is found \cite{Kuhn15b}.   Perhaps linear  morphologies (like DR~21 and NGC~2264) disperse rather than merge into simpler spherical morphologies (like W~40 and the three clusters of NGC~6357).  However, it seems physically reasonable to suggest that the dense but clumpy configuration of M~17 will equilibrate into a unified rich cluster.  

A second failure to detect (sub)cluster merging is from a scatter plot of total stellar population vs. $Age_{JX}$ for MYStIX (sub)clusters.  No indication of cluster population growth is seen \cite{Kuhn15b}.  It is possible that the statistical decomposition of stellar clustering into 142 isothermal ellipsoids masquerades a growth effect.  

A third test, however, gives a hint of cluster growth.  A strong anti-correlation between (sub)cluster central star densities and core radii  naturally appears in ensembles of young clusters.   A relationship $\rho \propto r_c^{-3}$ is expected from a collection of clusters of uniform and constant mass seen at different phases of expansion.  The MYStIX sample shows $\rho \propto r_c^{-2.6 \pm 0.1}$ over the range $0.03 \leq r_c \leq 1$~pc and $1.5 \leq \log \rho \leq 5$ stars/pc$^3$ \cite{Kuhn15b}.  This relationship appears shallower than a $-3$ powerlaw index, indicating  that larger clusters have somewhat higher masses than smaller clusters.  This suggests that MYStIX subclusters undergo growth from mergers or continued star fomation as they expand.  Note this this stands in contrast to Pfalzner's `leaky clussters' that lose mass as they expand \cite{Pfalzner09}.  

A fourth consideration gives a hint that merging may be needed to form the richest young clusters.  With the exception of W~3 Main \cite{Feigelson08}, there is no obvious case in the nearby Galaxy of a very rich (thousands of stars) with a dynamically relaxed appearance that is still embedded in its cloud.  The typical embedded cluster found in the MYStIX study has is not very rich (tens to hundreds of stars) and often with a clumpy morphology.  If rich clusters formed rapidly and monolithically as proposed by in some theoretical studies \cite{Elmegreen00, Palla00, Hartmann12}, then perhaps more should be found in embedded environments.  But a model where rich clusters form by the merging of smaller structures \cite{MacLow04, Bate09} is consistent with the paucity of very rich embedded clusters.  

\subsection{Duration of star formation}

The MYStIX and related studies give unequivocal evidence that long-lived star formation is pervasive, both across MSFRs and within rich clusters.  The acquisition of $Age_{JX}$ estimates for dozens of  spatially well-defined (sub)clusters allows us to study the history of star formation across MYStIX star forming regions.  Getman and colleagues find a clear and consistent pattern: more heavily absorbed clusters have younger ages than lightly absorbed clusters \cite{Getman14a}. This is shown for two MYStIX regions in Fig.~\ref{history.fig}, RCW~36 with a `simple' structure and Rosette Nebula with a `complex' structure. In RCW~36 the ages range from 0.9 to 1.9~Myr, while in Rosette they range from 1 to 4~Myr.   Ages are also available for  stars that are not assigned to clusters; these distributed stars always show older ages than absorbed clusters.  

These results confirm with widespread belief that clusters are formed inside dusty molecular cores (high $J-H$ color environments) and later expel their molecular material (low $J-H$ environments).  But there were few quantitative measures of this expectation prior to the MYStIX analysis.  Previous demonstrations of age gradients were based on spatial correlations between Class I-II-III (disk-bearing to disk-free) populations and absorption in the W~40 and Rosette Nebula regions \cite{Kuhn10, Ybarra13}.   Both of these quantities are not calibrated to age in Myr, and the situation is often not so simple; in the Orion L1541 cloud, for example, two clusters dominated by older disk-free stars are lightly obscured while one is heavily obscured \cite{Pilliteri13}.  
 
 \begin{figure}
\centering
\includegraphics[width=0.75\textwidth]{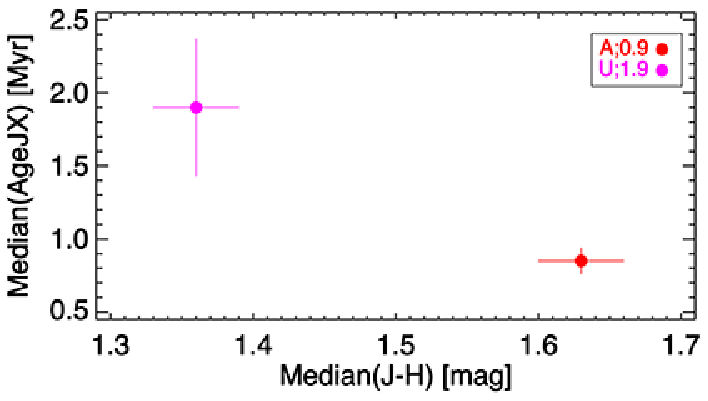} \\ \vspace{10pt}
\includegraphics[width=0.75\textwidth]{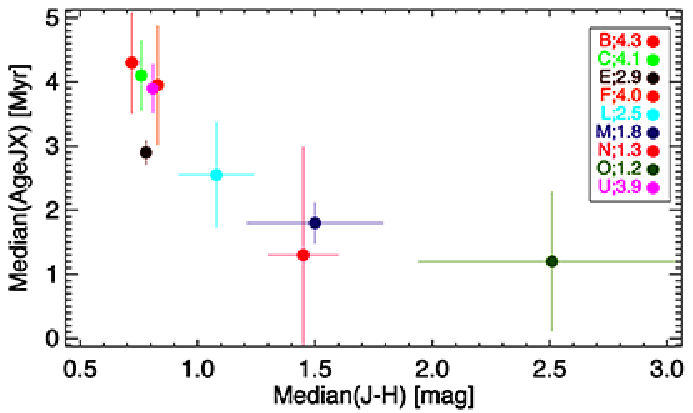}
\caption{Age differences for selected MYStIX subclusters as a function of $J-H$ color index, a measure of cloud absorption, in the RCW~36 and Rosette Nebula fields \cite{Getman14a}.  Each point refers to subclusters identified in \cite{Kuhn14a}.  The `U' designation refers to unclustered stars.  \label{history.fig}}
\end{figure}

\begin{figure}
\centering
\includegraphics[width=0.49\textwidth]{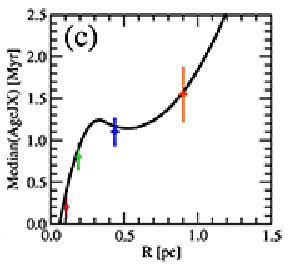}
\includegraphics[width=0.49\textwidth]{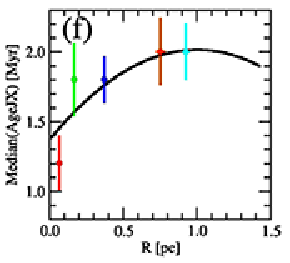}
\caption{Age differences within the Flame Nebula and Orion Nebula Clusters showing that the cores are younger than the halos \cite{Getman14b}. \label{corehalo.fig}}
\end{figure}

A more surprising result is the age spread, and spatial age gradient, found by  Getman and colleagues $within$ two nearby rich clusters, in addition to the gradients found earlier $between$ (sub)clusters \cite{Getman14b}. The cluster cores are much younger than the cluster outer regions (Fig.~\ref{corehalo.fig}).   In the Flame Nebula cluster,  stars within 0.2~pc of the center are 0.2~Myr old while stars 1~pc from the center are 1.6~Myr old.  In the Orion Nebula cluster, the age ranges from 1.2~Myr to 2.0~Myr.   This measurement is based entirely on analysis of solar-type stars, and thus does not conflate age and mass segregation.  

The result is startling because naive models for cluster formation (based on Jeans gravitational collapse in an isothermal cloud core) expect that star fill form first in the dense center, and thus would later appear to have the oldest, not the youngest stars.  Other models tend to homogenize the younger and older stars during a subcluster merging process \cite{Bate09}.  More complex cluster formation scenarios might explain the observed phenomenon;  for example, the older stars may have kinematically dispersed from the core, and/or the core may have been supplied with infalling molecular gas to allow star formation after the gas was depleted in the halo  \cite{Getman14b}.  

But the MYStIX intracluster age gradient also resolves a long-standing controversy concerning apparent stellar age spreads in HRDs \cite{Preibisch12}.  The age spread appears to be real, at least in part, because it represents a spatial segregation of older and younger stars.  Thus models based on rapid cluster formation in a single collapse time \cite{Elmegreen00} are not consistent with the findings, at least for the rich clusters in the Orion cloud complex. 

\section{Final comments and future reserach}

We emerge with some optimism that a frustrating period is ending when models for clustered star formation were largely unconstrained by empirical results concerning the outcomes of star formation processes (properties of the young stellar populations) to complement empirical results on the inputs to star formation processes (molecular cloud properties).  A multiwavelength approach provides the key: X-ray surveys to isolate the pre-main sequence population from the contaminating field star population and to avoid strong nebular emission; near-infrared imaging replacing optical observations to penetrate regions of high absorption; and mid-infrared photometry to discriminate the important subpopulation of disk-bearing young stars from often-overwhelming Galactic field star contamination. 

The diversity of clustering patterns found in MYStIX regions (Fig.~\ref{subcluster_morph.fig}) points to the importance of studying star formation in multiple environments.  The observational strategy of MYStIX can easily be extended to more star forming regions in the nearby (roughly distances $<3$~kpc) Galaxy.   Results are now emerging from Chandra X-ray Observatory observations of $\sim 20$ regions with distances $\leq 1$~kpc dominated by intermediate-mass BA stars \cite{Getman17}, and both Chandra and XMM-Newton missions have observed the nearest star forming regions around $0.14-0.3$~kpc.  

It is more difficult to extend such study to the lowest mass stars that dominate the IMF ($0 .1 < M < 0.5$~M$_\odot$), and to the richest star forming regions of the Galaxy lying $\sim 5-12$~kpc from the Sun.  Million-second Chandra exposures are needed to acquire sufficient X-ray sensitivity, and infrared followup requires both high resolution and high sensitivity.  Fortunately, the Chandra satellite is in good health since launch in 1999 and is likely to last for a considerable time into the future.  Infrared technologies are continuously improving:  the VISTA  Via Lactea project gives wide-field, multi-epoch photometry of large portions of the the Galactic Plane \cite{Minniti10}; the KMOS and MOSFIRE multi-object spectrographs offer efficient spectroscopic capabilities on 8-meter class telescopes; and the James Webb Space Telescope will greatly advance infrared imaging and spectroscopy in a few years.  These observational capabilities give confidence that fruitful interactions between theory and observations can become the norm in the study of clustered star formation.

\acknowledgement  This review rests on the labor and talents of the MYStIX team, particularly Patrick Broos, Konstantin Getman, Michael Kuhn, Tim Naylor, Matthew Povich, and Leisa Townsley.  Many of the astrophysical results appear in the dissertation of Michael Kuhn and work led by Kostantin Getman; the author is especially grateful for their collaborative energy and thoughtful analysis.  The MYStIX Project was principally supported at Penn State by NASA grant NNX09AC74G, NSF grant AST-0908038, and SAO/CXC grant AR7-18002X and ACIS Team contract SV-74018.

\end{document}